\documentclass[%
preprint,
 amsmath,amssymb,
 aps,
]{revtex4-1}
\usepackage{graphicx}
\usepackage{dcolumn}
\usepackage{bm}
\usepackage[margin=0.5in]{geometry}
\usepackage{amsmath}

\graphicspath{{/home/dino/Documents/Specific/Papers/OurPaper/Figures/}}

\begin{document}
\title{Chemically Active Nanodroplets in a Multi-Component Fluid}
\author{
Dino Osmanovi{\'{c}}
}
\affiliation{Center for the Physics of Living Systems, Department of Physics, Massachusetts Institute of Technology, Cambridge, MA 02139, USA}%
\author{
Yitzhak Rabin
}
\affiliation{Department of Physics, and Institute of Nanotechnology and Advanced Materials, \\Bar-Ilan University,
Ramat Gan 52900, Israel}%
\date{\today}
\begin{abstract}
We introduce a model of \textit{chemically active} particles of a multi-component fluid that can change their interactions with other particles depending on their state. Since such switching of interactions can only be maintained by the input of chemical energy, the system is inherently non-equilibrium. Focusing on a scenario where the equilibrium interactions would lead to condensation into a liquid droplet, and despite the relative simplicity of the interaction rules, these systems display a wealth of interesting and novel behaviors such as oscillations of droplet size and molecular sorting, and raise the possibility of spatio-temporal control of chemical reactions on the nanoscale.
\end{abstract}
\maketitle

\section{Introduction}

A particularly salient attribute of biological complexity is the highly multi-component nature of living cells. A normal cell contains thousands of different proteins which all exist in multiple copy numbers\cite{beck_2011} that can range over many orders of magnitude, even for the same protein in different cells\cite{maheshri_2007}.  Since live cells operate as chemical reactors, for a cell to be functional it is necessary for all of these proteins to find and bind their specific targets that can be other proteins, small molecules or specific DNA (RNA) sequences \cite{wodak_2013}.  These \textit{specific} interactions have to be realized in a sea of \textit{non-specific} interactions which can, in principle, interfere with the ability of proteins to find and bind their targets. Surprisingly,  the presence of non-specific interactions does not necessarily impede the formation of specific complexes; in fact, such ``non-functional'' interactions can help the diffusion-limited search by reducing the effective search space for specific partners through condensation and formation of membraneless compartments \cite{osmanovi_2016}. The realization that liquid-liquid phase separation on the nanoscale may play an important role in biology has received considerable attention recently \cite{brangwynne_2009,brangwynne_2011,hyman_2012,lin_2015,boeynaems_2018,elbaumgarfinkle_2015}. However, as is well known from from studies of microemulsions (e.g., oil droplets in water), in order to prevent coagulation and growth (Ostwald ripening) and eventual separation into macroscopic phases, one has to reduce the surface tension by adding surfactants or to maintain the microemulsion in a non-equilibrium steady state by the input of mechanical energy (stirring).  Another possibility that is perhaps more relevant to biology is that droplet size may be controlled by the input of chemical energy, e.g., through hydrolysis of ATP \cite{weber_2019}. A particular example of the latter has been explored by Zwicker et al \cite{zwicker_2016}, in what they deemed \textit{chemically active} droplets (see also \cite{seyboldt_2018}). By considering a system where a chemical energy source converts one kind of chemical species to another, with different solubility properties, they have observed life-like splitting of droplets. 

In order to gain qualitative insight about nanoscale separation in real biological systems, in this work we introduce and simulate a microscopic model of a multi-component fluid in which non-specific interactions between particles promote aggregation and formation of droplets while specific interactions  lead to the formation of bound pairs. Once a bound pair is formed, the interaction of its constituents with other particles changes (reminiscent of allosteric transitions in proteins following binding of ligands and ATP hydrolysis \cite{liu_2016,wodak_2019}) and this affects the stability of droplets. We find that such systems display new and unexpected non-equilibrium phenomena, such as  oscillations between states of aggregation (i.e., size and number of droplets) that depend on the number of different chemical components and the copy number of each component. We would like to emphasize that in this work we define as different chemical components as particles that interact only non-specifically with each other. Particles that can interact specifically with each other are deemed to be copy numbers of the same component. Our results point to the possibility of spatio-temporal control over chemical reactions in multi-component systems that are driven by input of chemical energy.

\section{Model}
Consider a dilute gas of $N$ particles (volume fraction $\eta\ll 1$) that consists of $N/M$ sets of $M$ particles each, where different sets represent chemically different species. All particles $i$ and $j$ within each set ($i,j\in M$) are identical in the sense that they interact with pair interaction parameter $\epsilon_{ij}=\epsilon_{S}$ (``specific" interaction), while particles $k$ and $l$ that belong to different sets ($k\in M, l\in M'$) interact with pair interaction parameter $\epsilon_{kl}=\epsilon_{NS}$ ("non-specific" interaction) such that $\epsilon_{S} \gg \epsilon_{NS}$. We assume that the interaction between particles $i$ and $j$ separated by a distance $r$ is given by the Lennard-Jones potential
\begin{equation}
\phi_{LJ}=4\epsilon_{ij}\left(\left(\frac{\sigma}{r}\right)^{12}-\left(\frac{\sigma}{r}\right)^6 \right)
\label{eq:LJ}
\end{equation}
where the particle size parameter $\sigma$ is assumed to be the same for all particles, and the interaction is cut off at $r_c>2.5$. When the distance between a pair of particles belonging to the same set is smaller than $1.5\sigma$ we say these particles form a "bound pair". Conversely, when this distance is $>1.7\sigma$  we refer to these complementary particles as a ``potential pair". We assume that {\it once a bound pair is formed, the interaction between each of the members of this pair and all other $N-2$ particles becomes purely repulsive} and is described by the Weeks-Anderson-Chandler potential $\epsilon_{NS}\left[4\left(\sigma/r\right)^{12}-1\right]$ for $r\le 2^{1/6}\sigma$ and zero otherwise \cite{weeks_1971}.  Since the energy associated with  non-specific interactions increases in the process, it must be supplied by an external source (e.g., by hydrolysis of ATP). Once a pair is formed, it has a characteristic lifetime and eventually decays into its constituent particles which interact with other particles in the system via the Lennard-Jones potential, Eq. \ref{eq:LJ} (with interaction strengths $\epsilon_S$ and $\epsilon_{NS}$ for particles in the same and in different sets, respectively).

\section{Results}\label{results}


 We begin our analysis with a system of $N=500$ particles and $M=2$, i.e., with 250 sets of 2 particles in each set. We choose the interaction parameters to be $\epsilon_S=6$ and $\epsilon_{NS}=2.2$  (this choice guarantees that the non-specific interactions are sufficiently strong to form liquid droplets \cite{osmanovi_2016}). We take $\eta=0.01$ since a low volume fraction is necessary in order to observe condensation into droplets. Though the only difference from our previous model is the change in the non-specific interactions once a bound pair is formed (this step requires input of chemical energy), the new model gives rise to strikingly different phenomena. 

\subsection{\textbf{Dynamic steady state}} The most striking feature in this system is that instead of the system evolving towards a single equilibrium state, it now settles into a dynamic steady state in which it oscillates between different states. These states are observed in figure \ref{fig:evo} and are summarized below (also see movie M1 in Supplementary Information):

\begin{figure}[h]
\begin{center}
\includegraphics[width=90mm]{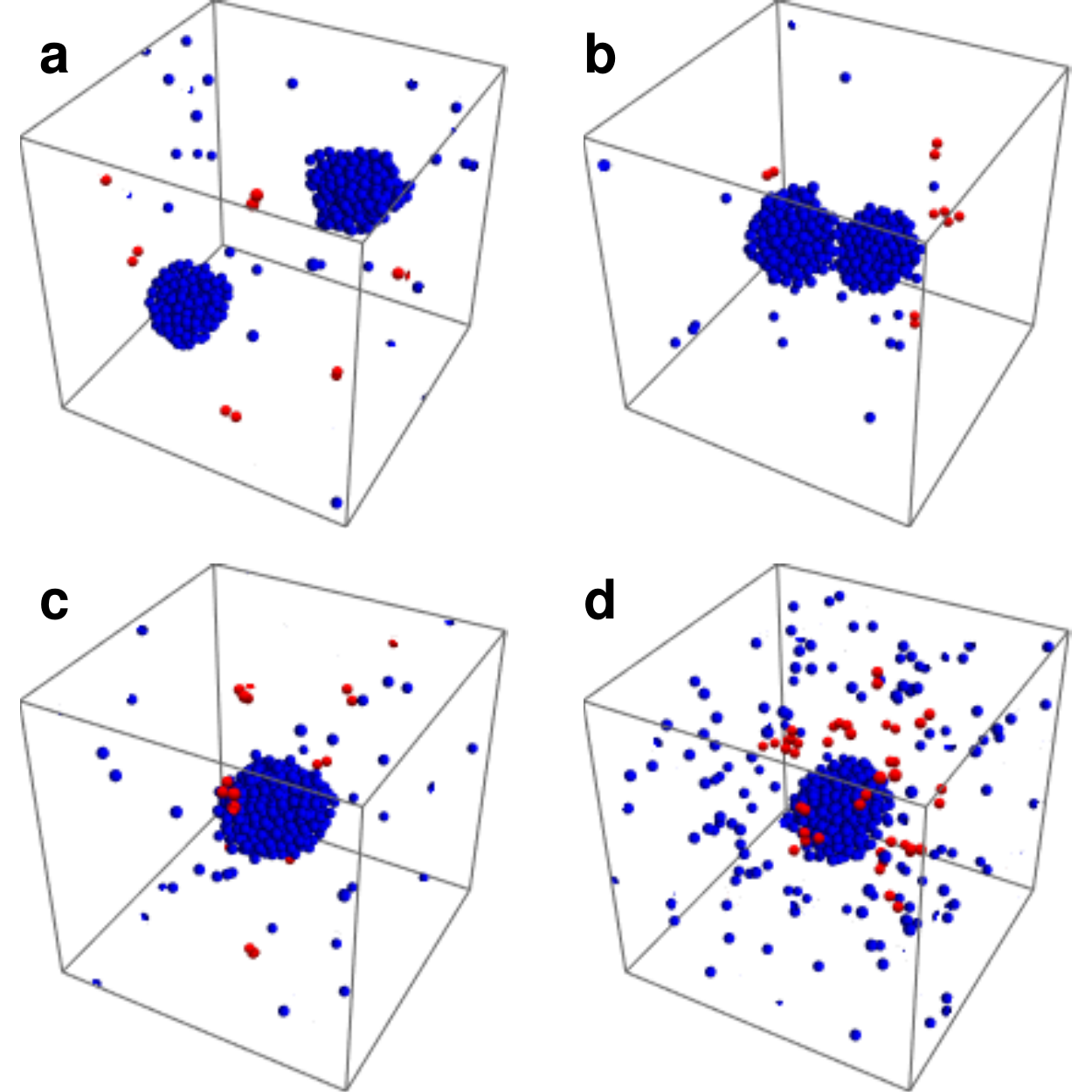}
\caption{The stages of droplet time evolution for a chemically active system. Particles are colored blue unless they are in a bound pair, in which case they are repainted red. The snapshots show the progress from the two-droplet state (a) the fusion of two droplets (b) the single-droplet state (c) and the decay of the large droplet through ejection of pairs (d).  The stages are then repeated again.}
\label{fig:evo} %
\end{center}
\end{figure}

\begin{itemize}
\item \textbf{Coexistence of two droplets} As can be seen from the figure, the system quickly separates into two clusters of roughly equal size, which coexist for relatively long periods of time.
\item \textbf{Fusion of droplets} The two separate droplets diffuse and, upon encountering each other, they undergo a fusion event leading to the formation of a large droplet which contains most of the particles in the system. 
\item \textbf{Decay of the large droplet} The fusion of the two smaller droplets is associated with increase in the number of bound pairs that form inside the large droplet. These bound pairs are rapidly ejected from the droplet leading to its fast decay.
\item \textbf{Nucleation of a second droplet} As the large droplet looses particles, it gradually decays until it contains about half of the particles in the system. Bound pairs dissociate in the gas phase and isolated particles aggregate and form another droplet of similar size. Eventually the first step is repeated. 
\end{itemize} 

\begin{figure}[h]
\begin{center}
\includegraphics[width=90mm]{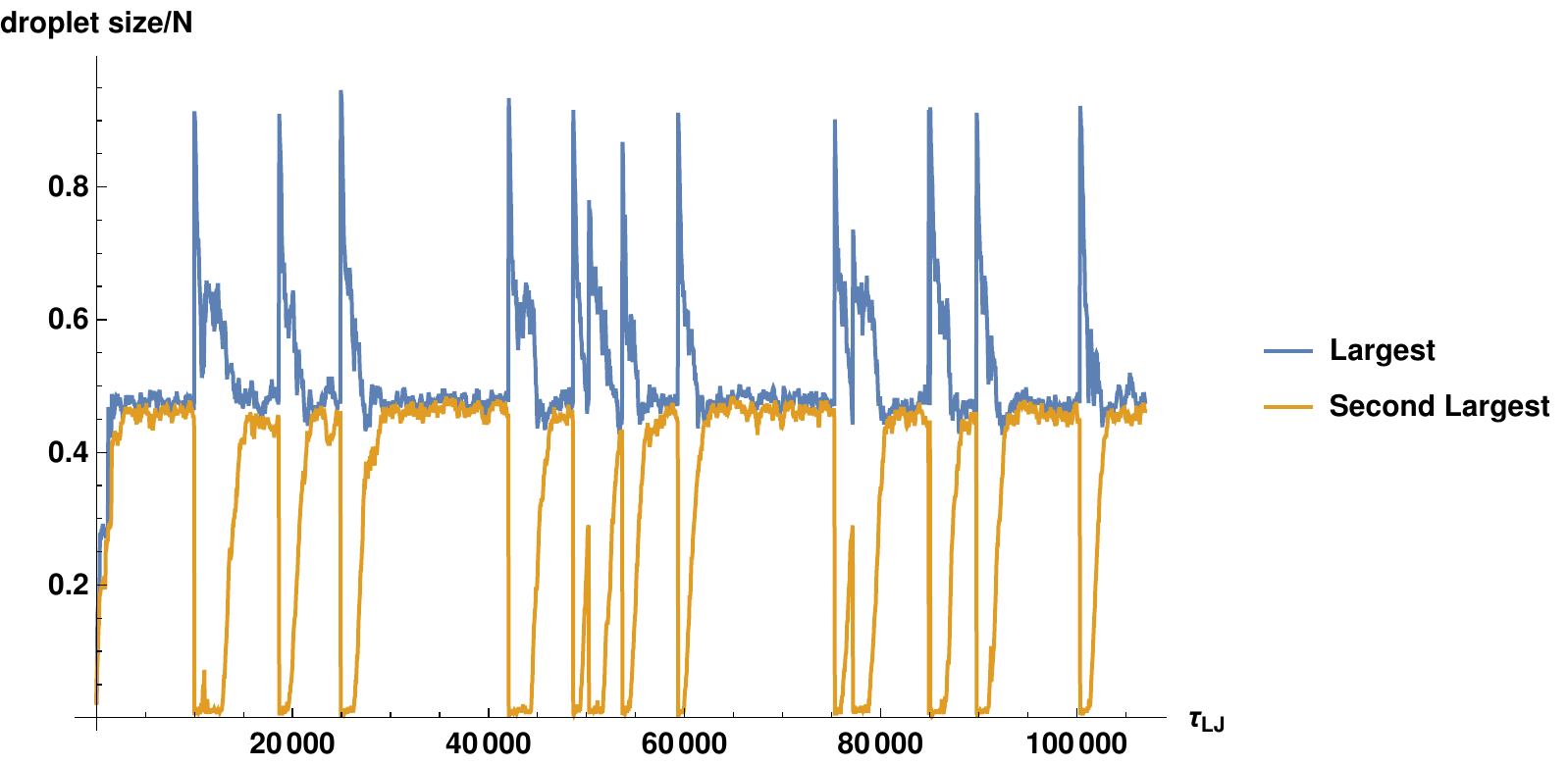}
\caption{The fraction of particles in the largest and the second largest droplets in the system is plotted as a function of time. }
\label{fig:tt} %
\end{center}
\end{figure}

The steps outlined above can be quantified by tracing the number of particles $N_c$ in the largest and the second largest droplets against time in Fig. \ref{fig:tt}. 

\subsection{\textbf{Pair Formation}}
\begin{figure}[h]
\begin{center}
\includegraphics[width=180mm]{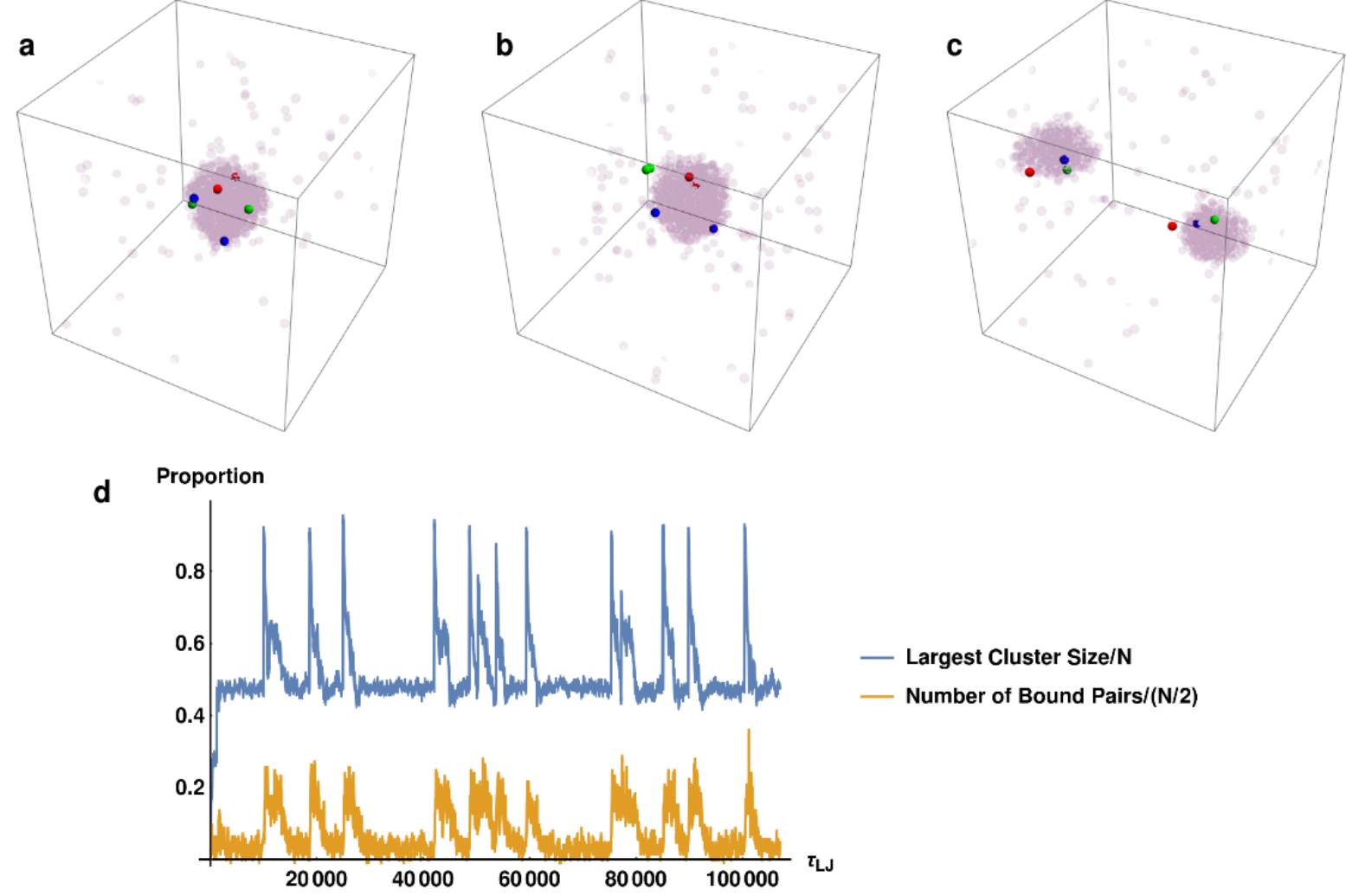}
\caption{Highlighting a few randomly chosen potential pairs as opaque particles of the same color, as the system evolves.(a) When the the droplet contains most of the particles in the system, all the potential pairs are inside the droplet. (b)  As they find each other and form bound pairs, they are rapidly expelled from the droplet.(c) Finally, the bound pairs decay and the system settles into a steady state where complementary particles are found in the different droplets.  (d) The fraction of particles in the largest cluster and the fraction of bound pairs in the system are plotted as a function of time. It can be seen that they are strongly correlated. }
\label{fig:pair} %
\end{center}
\end{figure}

In the previous subsection, we elucidated what happens to the system on a large scale. Now we will discuss the connection between droplet dynamics and the formation of specific complexes (bound pairs). In Figs. \ref{fig:pair}a-\ref{fig:pair}c we present several typical snapshots of the system, with 3 randomly chosen potential pairs (i.e., particles that interact specifically with each other and can form a bound pair upon close approach) shown in different colors. In Fig. \ref{fig:pair}d  we track the fraction of bound pairs (number of bound pairs divided by the total number of pairs $N/2=250$) and the fraction of particles in the largest droplet, as a function of time.
It can clearly be seen from figure \ref{fig:pair}d that the proportion of pairs is strongly correlated with the size of the largest droplet. The moment at which the two droplets fuse to form a larger droplet corresponds to an explosion in the number of bound pairs. These pairs then change their non-specific interactions with  the surrounding particles in the droplet from attractive to repulsive, causing them to be ejected from the droplet (see Fig. S1 in Supplementary Information). Formation of bound pairs is directly correlated with the decrease in the size of the droplet as pairs continue to be ejected from it. Ejected bound pairs dissociate in the gas phase outside the droplet, and the  released particles interact attractively with other particles in the gas. Eventually the ejected material nucleates a second droplet that grows until its size approaches that of the first droplet and the process repeats itself.

 \begin{figure}[h]
\begin{center}
\includegraphics[width=180mm]{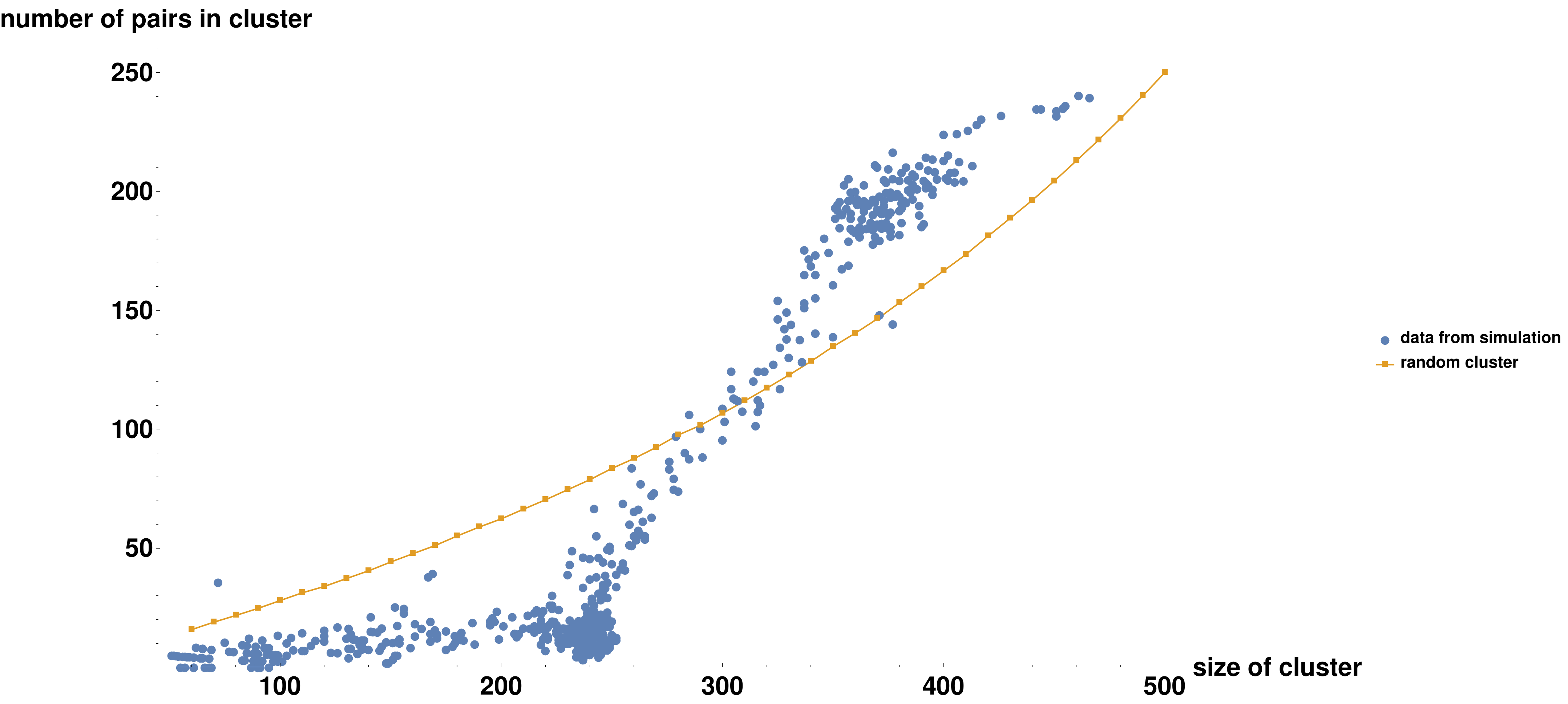}
\caption{Number of pairs in a droplet of size $N_c$ for the $N=500$, $M=2$ system, sampled at different times throughout the simulation. The orange line is what would be expected if the droplets were composed randomly. The blue points are calculated directly from clusters found in the simulation. }
\label{fig:tp} %
\end{center}
\end{figure}

In order to get additional insight into the interplay between droplet dynamics and formation of bound pairs, we consider the total number of particle pairs (both potential and bound pairs) $n_p$, contained in droplets of various sizes. Figure \ref{fig:tp} shows in a striking way how the simulation values differ from those calculated for a random system (a droplet made of randomly chosen particles). While in the random case the number of pairs increases roughly quadratically with $N_c$, the simulation results deviate from this curve. Thus, for droplets larger than about 300 there is overabundance of pairs compared to the random value, and for droplets smaller than 300 the number of pairs is below that for a randomly composed droplet.  In order to understand what is going on we first focus on the small clusters with $N_c<250$, in which the average number of pairs per cluster $n_p$ is very small and nearly independent of cluster size $N_c$.  Clearly, we are observing the phenomenon of \textit{active sorting} in which the only one of the two complementary particles of each set is found in the small cluster while the other particle is found either in another small cluster (or in the remaining gas phase). 
To understand this phenomenon note that as bound pairs are ejected from the decaying droplet, the number of pairs in this droplet decreases.  The emitted bound pairs dissociate in the gas phase and, eventually, another droplet is nucleated somewhere in the system. This droplet grows mostly by absorbing single particles, with a tendency to eject any bound pair that forms. Such pairs break up in the gas phase and their constituents are reabsorbed by the two droplets. This leads to effective ``sorting" where the two complementary particles of each set have the tendency to be found in different droplets, each of which is roughly of  half the system size. This explains why the number of the potential pairs is so much lower than the random result for droplets of size $N_C<250$, and also provides intuition for why the size of each of the two coexisting droplets is stabilized somewhat below $N_C=250$ (recall that some particles remain in the gas phase). Note that the condition for this mechanism to be valid is that the diffusion controlled encounter time between two members of the pair in a droplet should be much shorter than the time between the nucleation of the second droplet and the coalescence of the two droplets. This condition is indeed satisfied in our simulation (see table 1 for the relevant timescales in this simulation).

Now lets return to Fig. \ref{fig:tp} and consider large droplets of size $N_c>250$. Clearly such droplets correspond to the time intervals during which a single large droplet exists (see Tab. \ref{fig:tt}). Since this large droplet is formed by coalescence of two small clusters that are enriched in complementary members of all pairs, the average number of potential pairs within it exceeds that expected for a randomly composed droplet. The broad crossover region corresponds to  time intervals (Fig. \ref{fig:tt}) in which a single large droplet decays by emission of bound pairs.
\begin{table}[h]
\begin{center}
\includegraphics[width=90mm]{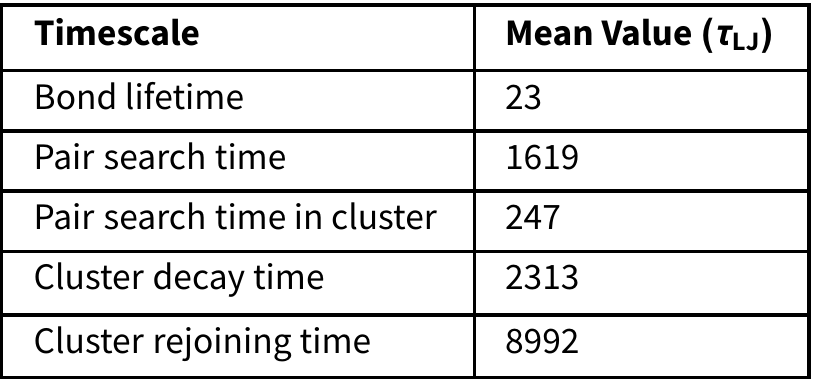}
\caption{Different time scales for the $N=500$, $M=2$ system. }
\label{fig:timescales} %
\end{center}
\end{table}

\subsection{Large sets} \label{sec:larg}

\begin{figure}[h]
\begin{center}
\includegraphics[width=180mm]{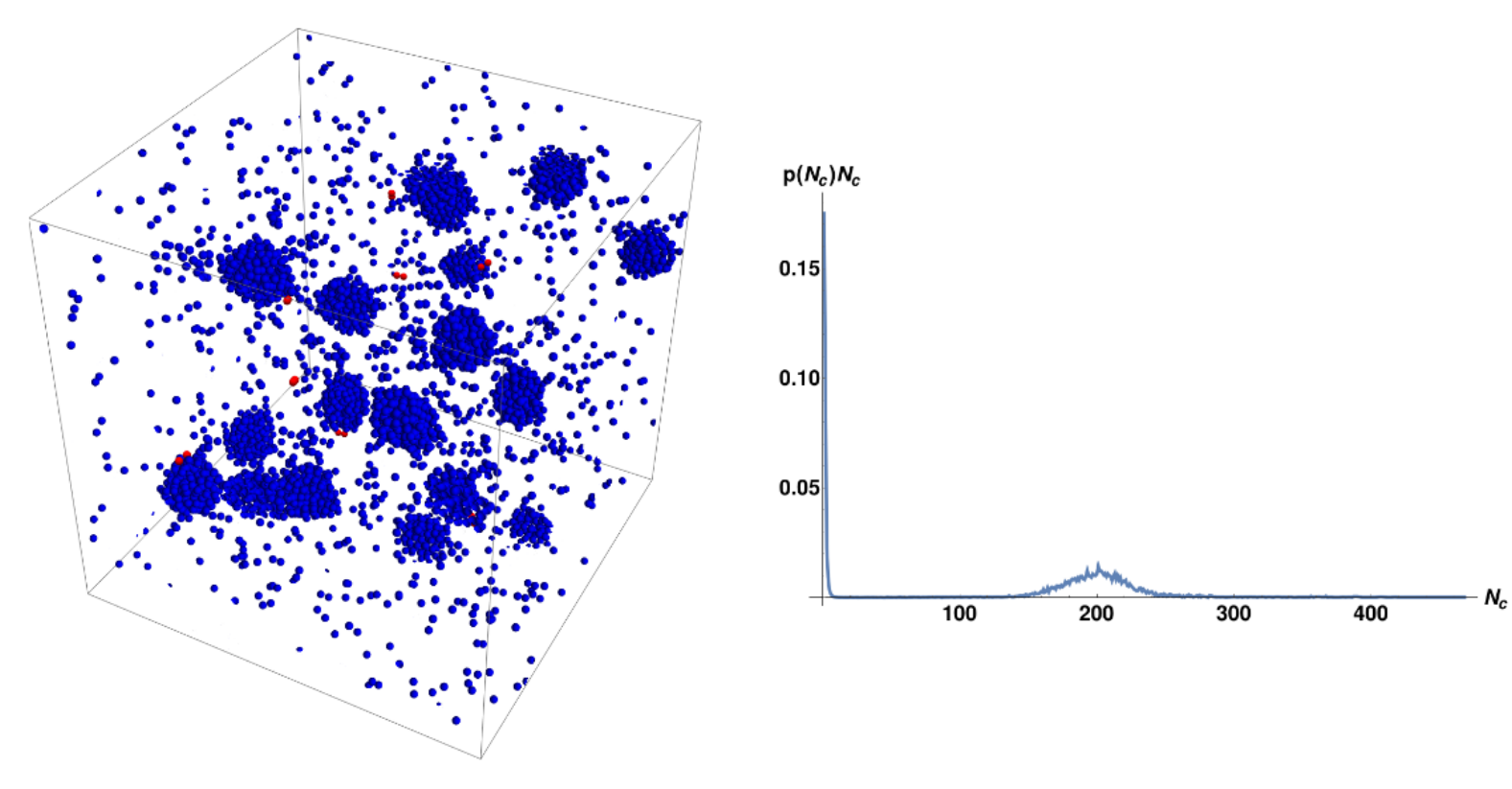}
\caption{ a) A snapshot of a chemically active system with $N=3750$ and $M=15$. The system undergoes microphase separation into many droplets of approximately equal size. b) The normalized droplet size distribution multiplied by droplet size.}
\label{fig:bigsys} %
\end{center}
\end{figure}

We now consider what occurs as we increase the number of particles per set $M$ (recall that all $M$ particles within a set can form bound pairs with each other) while keeping the same number of sets ($250$) as in the previous sections. This is a more realistic representation of a chemically heterogeneous fluid where multiple particles can interact specifically with each other. The interaction rules are the same as presented in the ``Model" section. The behavior of a system with with $M=15$ and $N=3750$  can be seen in Fig. \ref{fig:bigsys}.

Inspection of Fig. \ref{fig:bigsys} shows that in the larger system there are many droplets rather than just the two observed in the smaller system. The average size of a droplet in this system is slightly lower than in the $M=2$ case. Beyond this the droplets have similar dynamics including coalescence events of two droplets, followed by decay of the resulting droplet by emission of bound pairs (see movie M2 in Supplementary Information), as seen in the previous sections. 


 \begin{figure}[h]
\begin{center}
\includegraphics[width=180mm]{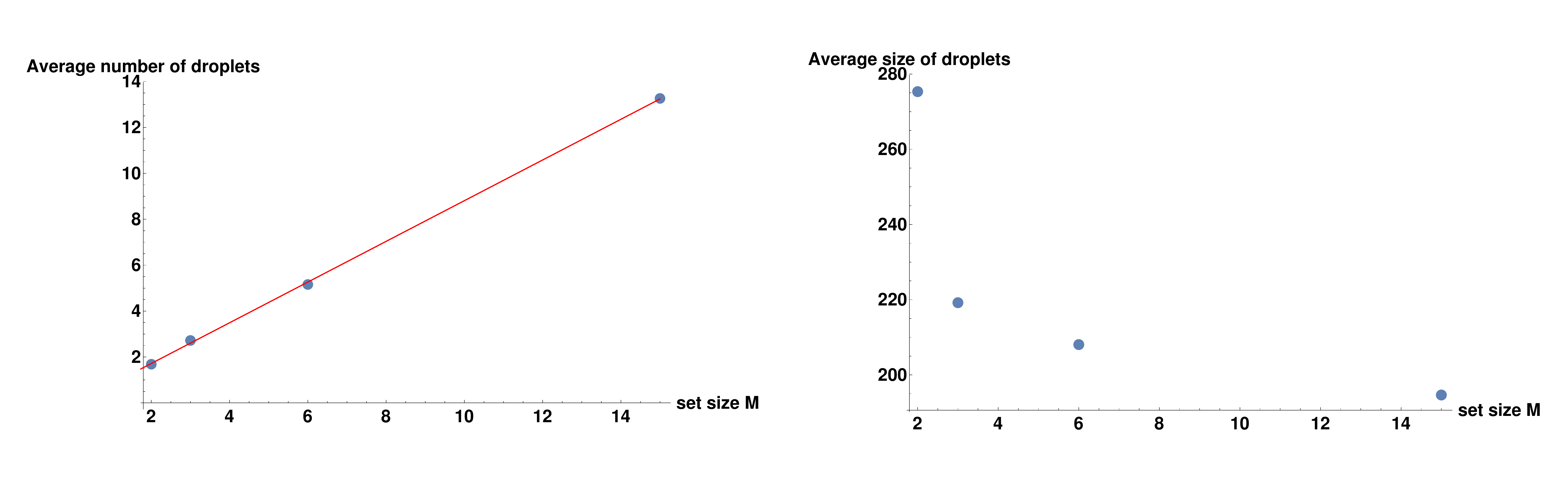}
\caption{ a) The total number of droplets is plotted for different set sizes $M$. The red line shows the best linear fit corresponding to a curve of 0.88*M. b) The average size of droplet as a function of the set size $M$.}
\label{fig:sysscale} %
\end{center}
\end{figure}

We proceed to explore the full range of behaviors as we keep the number of sets fixed and change the amount of particles per set $M$. We plot the trends for set sizes of $M=2,3,6,15$ in Fig. \ref{fig:sysscale}. As can be seen in Fig. \ref{fig:sysscale}, the number of droplets increases linearly with the set size. From these plots we can identify a rule of thumb for these chemically active, multi-copy systems, both in terms of the number and the average size of droplets. It would appear that the number of droplets is approximately equal to the size of the set and the average size of the droplet is approximately equal to the total number of sets $N/M$. This accords with our previous observations that the members of a set will tend to sort themselves into different droplets. Obviously this rule of thumb does not entirely rule true and deviations from it can be understood by considering how many of all the possible pairings in the system ($N(N-1)/2$) can form bound pairs, ($(N/M)M(M-1)/2)$). Since the fraction of pairs goes as $(M-1)/(N-1)$, it grows when we keep $N/M$ fixed while increasing M. This means that the ``latent" proportion of possible specific chemical bonds increases as the set size M increases (though it saturates quickly as M becomes large). This increased probability of encountering a member of the set means that droplets are slightly smaller, up to the saturation threshold.

\section{Discussion}

While the model we have presented is fairly simple, it captures a range of complex behaviors that arise from the fact that non-specific interactions  between particles can change as the result of binding between specific partners. In this section we go over some of the consequences that could have some relevance for real systems.

\subsection{Non-equilibrium chemical reactions}
The way the interaction rules have been implemented defines the system as a non-equilibrium one. The changing interaction of the non-specific bonds requires an implicit input of energy in order to form specific bonds. In the simulation, this energy cost is effectively paid for immediately by removing the attractive part of the non-specific interactions upon the length of the specific bond being less than $1.5\sigma$, thus increasing the total energy of the system. We would here like to comment on several aspects of this and how it relates to real systems.

The nature of the defined interaction rules can be understood as a particular form of a three body potential $V({\bf r_1},{\bf r_2},{\bf r_3})$, by this we mean that the state of randomly chosen particles $1$ and $2$ in a set (their separation $|{\bf r_1}-{\bf r_2}|$) will affect their interaction with any other particle $3$. A more realistic formulation of this three body potential would introduce a barrier for the formation of a specific bond were the particles under consideration already bound non-specifically. The difference in energy between the situation where there is a specific bond and that where the particles are bound to non-specific neighbors would have to be either supplied from the bath or by an non-equilibrium energy input. In addition to this, the traversal over the barrier could be helped by catalysts present in the system. 

An alternative scenario which would have similar physics but be more challenging to simulate would involve the energy of the specific interactions being orders of magnitude larger than the non-specific interaction. In this case the formation of specific bonds would be very energetically favored, and they would also be expelled from the droplet. However, in such a system the thermal energy would not be sufficient to dissociate the specific bonds, thus favoring the formation of a diatomic gas. Yet, if we were to provide an energy input in the gas phase to break apart the specific bonds, we would anticipate seeing similar results to the current work. In order to put the present work in a proper context we would like to mention that our system belongs to a broader class of chemical reactions that give rise to non-equilibrium spatio- temporal patterning (see, e.g., the Belousov-Zhabotinskii reaction \cite{murray_1993}).

\subsection{Storage and production modes}

Of particular interest is the spatio-temporal organization of the multitude of different chemical components in the system. It is apparent from the results that combination of (a) weak non-specific interactions between particles in different sets that leads to phase separation and formation of droplets (b) bound pair formation due to strong attractive interaction between particles within a set that changes the interaction between the constituents of these pairs and all other particles to repulsive and (c) the dissociation of these bound pairs, results in oscillations between ``storage'' and ``production'' modes. In the storage phase, there is effective sorting (segregation) of the different members of each set into different droplets that act as "storage tanks" whose number and size depends on the size and number of sets $M$ and $N/M$, respectively. The number of complexes (bound pairs) formed in the storage phase is low since complex formation is greatly suppressed through members of the set being found in different droplets.  The duration of the storage phase depends on the concentration of droplets that increases with increasing set size $M$. Coalescence of these droplets gives rise to the appearance of larger droplets that act as ``chemical reactors" in which complementary members of the sets present combine to form bound pairs, a production mode that requires the input of chemical energy. These bound pairs are then ejected from the large droplets and diffuse in the surrounding gas phase until they finally dissociate (the energy released in the process is dissipated) and are absorbed by the droplets, and the system returns to the storage phase.  The duration of the period between production and storage phases increases with the lifetime of the bound pairs and it is during this intermediate phase that metastable complexes exist in the gas phase and may play a functional role (e.g., as catalysts/enzymes) in biological or in artificially engineered systems.

As one goes to larger, more realistic, systems, the same features are present but there is no longer a global temporal separation between production and storage. Instead droplets exist relatively stably until they encounter another droplet, in the same way as is seen in the smaller system, but now there are many droplets in the system. Therefore the production of complexes is tied to the dynamics of all the droplets.


\subsection{Nanoemulsions of chemically heterogeneous fluid systems} 

Phase separation has been a topic of recent interest in biological systems\cite{boeynaems_2018}. However, some of its observed features suggest that such phase separated systems are not at equilibrium. Recall that a normal liquid-liqiud phase separation leads to macroscopically segregated phases, rather than to many small droplets immersed in the surrounding medium (the classic oil and water example would have small droplets of oil in upon immediate mixing, but were the system to be left for a while the droplets would coalesce and two macroscopically separated phases would be observed). The current scheme provides a microscopic basis for how phase separation on the nanoscale might occur in a fluid, the necessary requirements being that the fluid is chemically heterogeneous and that the microscopic interactions can change upon binding. 

The system selects a length scale of the droplets based on the degree of chemical heterogeneity. As can be seen in section \ref{sec:larg} the degree of chemical heterogeneity, represented in our case by the number of different sets $N/M$, will determine the subsequent size of the nanodroplets. While the precise mechanism in real systems might differ in details, especially with regards to the interactions within sets, it is clear that microphase separation requires a way to suppress Ostwald ripening. In the case studied in the present work, droplets above a certain size have a tendency to form bound pairs that are expelled from the droplet, thus suppressing the formation of very large droplets.

\section{Conclusions}\label{conclusions}

In this paper we have explored the behavior of a chemically heterogeneous system in which the interactions between various particles in the system depends on the particular spatial configuration of the particles. In particular, we introduce a scheme whereby the system is divided into sets such that when any two members of a set encounter each other, they change their interactions with all the other particles in the system. This scheme was introduced to mimic real chemical bonds, but many different schemes where there are state-dependent interactions also exist in a variety of settings, leading to potentially many differing kinds of behaviors. For example, the attractive interaction might be simply reduced instead of completely switched off, or there might be an energetic barrier to binding. One could also imagine scenarios where the set size was different for different sets.

We have observed a number of physically interesting and potentially biologically-relevant phenomena. The system is maintained in a non-equilibrium state by the input of chemical energy required to change the way in which particles that form bound pairs, interact non-specifically with other particles in the system. Despite the fact that the energy input is on the molecular scale (change of the local interactions), its effects are manifested on a much larger scale than this. One immediate observation is that very large droplets become unstable because they contain too many particles which can form bound pairs that are ejected from the droplet. This naturally leads to the formation of smaller droplets in which the number of particles that can form bound pairs is drastically reduced. Of course, such smaller droplets are unstable with respect to coalescence and therefore a dynamic steady state in which the system oscillates between larger and smaller droplet states, is established. While this droplet dynamics is the most striking large-scale feature of the system, other facets of its behavior become apparent if we focus our attention on the chemical composition of the droplets. In the regimes studied here the molecules that form specific bonds have a tendency to sort themselves into different droplets. Therefore the system becomes organized not only in the sense that multiple liquid droplets of similar size form, but also in terms of the highly non-random chemical makeup of those droplets. This makes these smaller droplets quasi-stable, but upon encountering another droplet, the high degree of chemical compatibility will lead to an explosion in production of chemical compounds (bound pairs), which are immediately expelled from the larger droplet. This is a possible method of controlling the output of chemical compounds on the nanoscale in real systems. Whether similar mechanisms play a role in the functioning of living cells, remains an open question.

\section{Conflicts of Interest}

The authors declare no conflicts of interest

\section{Acknowledgements}
DO would like to thank the Gordon and Betty Moore foundation for funding. YR's work was supported by grants from the Israel Science Foundation and from the Israeli Centers for Research Excellance program of the Planning and Budgeting Committee. 

\bibliographystyle{unsrt}
\bibliography{exported-references}

\end{document}